\pdfoutput=1

\documentclass[aps,prd,onecolumn,groupedaddress,preprintnumbers,superscriptaddress,nofootinbib]{revtex4}
\usepackage{amsmath,amssymb,latexsym}

\begin{document}

\title{
Effects of Light Fields During Inflation
}


\author{
Takeshi Kobayashi
}\email[]{takeshi.kobayashi@ipmu.jp}
\affiliation{
Institute for Cosmic Ray Research, The University of Tokyo, 5-1-5
Kashiwanoha, Kashiwa, Chiba 277-8582, Japan 
}
\author{
Shinji Mukohyama
}\email[]{shinji.mukohyama@ipmu.jp}
\affiliation{
Institute for the Physics and Mathematics of the Universe (IPMU), The
University of Tokyo, 5-1-5 Kashiwanoha, Kashiwa, Chiba 277-8583, Japan
}
%\thanks{}

\preprint{ICRR-Report-561} 
\preprint{IPMU10-0038}

%\date{\today}

\begin{abstract}
In the inflationary universe, there can be light fields other than the
 inflaton. We explore a possibility that such light fields source the
 primordial perturbations, while minimally affecting the inflaton
 dynamics. We show that during inflation, fluctuations of the light
 fields can be converted to adiabatic curvature perturbations, which
 accumulate and become significant by the end of the inflationary
 era. An additional goal of this work is to distinguish between light
 fields which can/cannot be ignored during inflation. Such criteria
 become useful for examining cosmological scenarios with multiple
 fields. As concrete examples, our results are applied to D-brane
 inflation models. We consider effects from KK modes (oscillation modes)
 of wrapped branes in monodromy-driven large-field models, and angular
 directions of throat geometries in warped D-brane inflation.
\end{abstract}

%\pacs{}
\maketitle

\section{Introduction}
\label{sec:intro}

The initial conditions for the Hot Big Bang cosmology can be set by
having an exponential expansion phase in the early
universe~\cite{Starobinsky:1980te,Sato:1980yn,Guth:1980zm}. A simple way
to drive such an inflationary stage is to have a scalar field, dubbed
the inflaton, whose potential energy dominates the universe. However,
microscopic descriptions of inflationary cosmology in many cases
predict, or even require the existences of additional light fields
during this era. One may expect that such light fields show up merely as
tiny corrections to the inflaton action, having negligible influences on
the inflaton dynamics. In this work, we show that even in such cases,
light fields can leave significant imprints on the primordial curvature
perturbations. We explore the possibility that light fields other than
the inflaton generate the dominant contributions to the curvature
perturbations while minimally affecting the inflaton
dynamics. This may open up new possibilities for inflationary models in
which the inflaton field cannot produce perturbations compatible with
observational data. 

An additional goal of this work is to distinguish between light fields
which can and cannot be ignored during inflation.
This is related to questions such as: When can we
treat multi-field inflation models as single-field ones? What are the
effects of unstabilized moduli during inflation? 
Implications of light fields on cosmology after inflation have been
discussed extensively as the moduli
problem~\cite{Coughlan:1983ci,Banks:1993en,de Carlos:1993jw}, while in 
this paper we discuss the effects of light fields during inflation,
especially on the generated curvature perturbations.
We discuss conditions under which the existence of light fields
can/cannot be ignored. This may have relevance to mechanisms such as 
curvaton~\cite{Linde:1996gt,Enqvist:2001zp,Lyth:2001nq,Moroi:2001ct} and
modulated reheating~\cite{Dvali:2003em,Kofman:2003nx} scenarios which
require (and often simply assume) light fields to have no
effects during inflation.  

How light fields generate curvature perturbations can simply be
understood as follows: When a light field~$\sigma$ shows up as
corrections to the inflaton action, then the number of
e-foldings~$d\mathcal{N}$ obtained within a certain inflaton field
range~$d\phi$ also becomes dependent on $\sigma$,
$d\mathcal{N} = F(\phi, \sigma)\, d\phi$. Therefore it is clear
that fluctuations of $\sigma$ lead to inhomogeneous expansion of the
universe, generating cosmological perturbations. 

One can also consider the light field as one of the inflatons, and
interpret the system as a multi-field inflation model. Then the
perturbations can be understood as arising from patches of the
universe taking different paths in field space, where the entropy
perturbations are converted to curvature perturbations during
inflation (see
e.g.~\cite{Polarski:1992dq,GarciaBellido:1995qq,Gordon:2000hv,GrootNibbelink:2001qt,Sasaki:2008uc}). 
We will see that such perturbations originating from the fluctuations of
$\sigma$ can become dominant over the curvature perturbations from
$\phi$. 

As microscopic examples, our results are applied to D-brane
inflation models in string theory. We discuss the model proposed
in~\cite{Silverstein:2008sg} where monodromy of wrapped branes allows
large-field inflation. In this paper we further analyze effects of KK
modes (oscillation modes) of the wrapped branes which can become light
during inflation. We also comment on D-brane inflation in
a warped throat geometry~\cite{Kachru:2003sx} with light angular
directions. 

The rest of the paper is divided as follows. 
We first study effects of light fields when they correct the inflaton
action in the form~(\ref{L}) in Section~\ref{sec:II}. Then in
Section~\ref{sec:III}, we further discuss cases where the light field
corrections show up in different ways, as in (\ref{Lf}) and
(\ref{Lh}). The results are applied to D-brane inflation models in
Section~\ref{sec:IV}, where the details of the calculations are shown in
the appendix. We present our conclusions in Section~\ref{sec:conc}.

\section{Inflation Modulated by Light Fields}
\label{sec:II}

In this paper we study
implications of having additional light fields manifesting themselves as
corrections to the inflaton action. For this purpose, let us start by
considering the following toy Lagrangian,
\begin{equation}
 \frac{\mathcal{L}}{\sqrt{-g}} = -\frac{1}{2} g^{\mu\nu} \partial_{\mu}
  \phi \partial_{\nu} \phi \left(1 - f \frac{\sigma^2}{\mu^{2-m}
			    \phi^m}\right) - \frac{1}{2} g^{\mu\nu}
  \partial_{\mu} \sigma \partial_{\nu} \sigma - V(\phi) 
 \left( 1 + g \frac{\sigma^2}{\mu^{2-m} \phi^m}\right), 
\label{L}
\end{equation}
where the inflaton ($\phi$) kinetic term and potential are modified
by modulus ($\sigma$) corrections. ($\sigma$ can also be called the
inflaton, but for convenience we call $\phi$ the inflaton and $\sigma$
the modulus in this paper.) Though we have omitted the Einstein-Hilbert
term, we are assuming Einstein gravity.
The modulus corrections are suppressed by some mass scale~$\mu$, and
the field value of the inflaton~$\phi$ can also suppress the corrections
during inflation. $f$ and $g$ are dimensionless constants. Since we
consider cases where the modulus corrections are small, correction terms
that are higher order in $\sigma$ are neglected. 
Lagrangians of this type show up in many
cases, e.g., when having nonminimally coupled fields in the Jordan
frame, from nonminimal K\"ahler potentials in supersymmetric models. 
Later in Section~\ref{sec:IV} we give explicit examples in the context
of D-brane inflation models, where the moduli corrections are
suppressed by super-Planckian~$\phi$ (Subsection~\ref{subsec:KK}) or by
approximate symmetries (Subsection~\ref{subsec:angular}), thus allowing
the moduli to be light during inflation.
One may encounter microscopic models
involving more complicated moduli corrections, but the crucial points
can be captured by studying the above Lagrangian. Cases where the
$\sigma$-corrections to the inflaton kinetic term and potential have
different minima are later studied in Subsection~\ref{subsec:h}.

We discuss curvature perturbations arising from fluctuations~$\delta
\phi$ and $\delta \sigma$. Since we focus on cases where the 
$\sigma$-corrections to the inflaton action are small, $\delta \phi$
almost directly generates adiabatic curvature perturbations at the time
when the fluctuations exit the horizon. On the 
other hand, $\delta \sigma$ mainly produces entropy perturbations which
can be converted to curvature perturbations during inflation. This 
kind of curvature perturbations arise due to 
$\delta \sigma$ kicking patches of the universe to different classical
trajectories in the $\phi - \sigma$ plane, generating perturbations in
the number of e-foldings~$\delta \mathcal{N}$ among different trajectories. 
As we will soon see, when the entropy perturbations from $\delta
\sigma$ are sufficiently transformed to adiabatic ones, then $\partial
\mathcal{N}/\partial \sigma$ becomes dominant over $\partial
\mathcal{N}/\partial \phi$.  

The modulus correction to the inflaton kinetic term (i.e. the term
with the coefficient~$f$ in (\ref{L})) modifies the inflaton velocity,
whereas the correction to the inflaton potential (i.e. the term with
$g$) modifies the inflaton velocity as well as the Hubble parameter
during inflation, cf. (\ref{11}) and (\ref{12}).
Therefore inflation proceeds differently among
different trajectories in the $\phi -\sigma$ plane.
$\delta \mathcal{N}$ produced in this way is amplified when the
trajectories during inflation diverge and differences among patches
expand. On the other hand, in cases where the trajectories converge, the
later transformation of the entropy to adiabatic perturbations is
suppressed.\footnote{More precisely, distances among
trajectories in field space do not necessarily reflect how differently
inflation proceeds among patches.
One should keep in mind that the 
diverging/converging discussion in the text is adopted for illustrative
purposes.} In this paper, no matter the trajectories diverge or converge
during inflation, we assume that after the inflationary era the
trajectories converge to a single one (as is the case for the examples in
Section~\ref{sec:IV}). \\ 

Now let us actually compute $\delta \mathcal{N}$ produced from $\delta
\phi$ and $\delta \sigma$, and see under which conditions the curvature
perturbations sourced by the modulus field become significant. 

In a flat FRW universe, the equations of motion for $\phi$ and $\sigma$,
and the Friedmann equation are
\begin{equation}
  \ddot{\phi} \left(1 - f \frac{\sigma^2}{\mu^{2-m} \phi^m}\right) 
 + 3 H \dot{\phi} \left\{ 1- \left( 1 + \frac{2}{3}
			      \frac{\dot{\sigma}}{H \sigma } -
			      \frac{m}{6} \frac{\dot{\phi}}{H
			      \phi}\right) f\frac{\sigma^2}{\mu^{2-m}
 \phi^m}\right\}  
 = - V' \left\{ 1 + \left(1 - \frac{mV}{V'\phi}\right) g
       \frac{\sigma^2}{\mu^{2-m} \phi^m}\right\}, 
\label{1} 
\end{equation}
\begin{equation}
 \ddot{\sigma} + 3 H \dot{\sigma} = - (2 g V + f \dot{\phi}^2) 
 \frac{\sigma}{\mu^{2-m} \phi^m}, 
\label{2}
\end{equation}
\begin{equation}
 3 H^2 M_p^2 = \frac{1}{2} \dot{\phi}^2 \left( 1 - f 
 \frac{\sigma^2}{\mu^{2-m} \phi^m}\right) + \frac{1}{2} \dot{\sigma}^2 + 
 V \left( 1 + g \frac{\sigma^2}{\mu^{2-m} \phi^m}\right),
\label{3}
\end{equation}
where derivatives with respect to $\phi$ are denoted by primes, and
time~$t$ derivatives by overdots. 

We assume in this paper that $\phi$ is under slow-roll inflation, though
estimations can be carried out in a similar manner with inflation models
giving different inflaton dynamics,
e.g. rapid-roll~\cite{Linde:2001ae,Kofman:2007tr,Kobayashi:2009nv},
Dirac-Born-Infeld (DBI) inflation
models~\cite{Silverstein:2003hf,Alishahiha:2004eh}. Also, we focus on
cases where the $\sigma$-corrections to the inflationary dynamics are
tiny, therefore, (\ref{1}) and (\ref{3}) are approximated by 
\begin{equation}
 3 H \dot{\phi} \sim -V', 
\qquad
 3 H^2 Mp^2 \sim V. \label{slow-roll}
\end{equation}
We further assume 
\begin{equation}
 |g| \gtrsim |f|
\end{equation}
(cases with $g=0$ are discussed in Subsection~\ref{subsec:gzero}),
and that $\sigma$ is slow-rolling under the approximation
\begin{equation}
 3 H \dot{\sigma} \sim -2gV \frac{\sigma}{\mu^{2-m} \phi^m}.
 \label{sigma-g}
\end{equation}

Then one can check that the necessary conditions for the approximations
(\ref{slow-roll}) and (\ref{sigma-g}) to hold are the absolute values of
the following quantities be sufficiently smaller than one,
\begin{equation} \label{6}
\begin{split}
 &\epsilon \equiv \frac{M_p^2}{2}\left(\frac{V'}{V}\right)^2,
 \qquad 
 \eta \equiv M_p^2 \frac{V''}{V}, \\
 & g \frac{\sigma^2}{\mu^{2-m} \phi^m}, \qquad
 \frac{m V}{V' \phi} g \frac{\sigma^2}{\mu^{2-m} \phi^m}, \\
 & m\frac{\dot{\phi}}{H \phi} \sim -m \frac{M_p^2
 V'}{\phi V}, \qquad
 g \frac{M_p^2}{\mu^{2-m} \phi^m},
\end{split}
\end{equation}
where roughly speaking, the first line is the slow-roll conditions for
$\phi$, the second line for the modulus corrections to be minimal, and
the third line for slow-rolling of $\sigma$. The smallness of the last
quantity in the third line is equivalent to the effective mass of
$\sigma$ being smaller than the Hubble parameter, i.e. $m_{\sigma}^2 \ll
H^2$. \\

To make the discussion concrete, henceforth we consider
large-field inflation with
\begin{equation}
 V(\phi) \propto \phi^n,\quad \mathrm{where}\quad n = \mathcal{O}(1).
 \label{phin}
\end{equation}
Computations with other inflaton potentials are straightforward, though
we expect results obtained from (\ref{phin}) to be rather
general. Implications for other potentials including small-field
models will be briefly discussed at the end of the section.
Then, given $|m| \leq \mathcal{O}(1)$, the condition~(\ref{6}) is
translated to
\begin{equation}
 \frac{M_p^2}{\phi^2}\ll 1 ,\qquad 
 \left| g \frac{\sigma^2}{\mu^{2-m} \phi^m}\right| \ll 1,\qquad
 \left| g \frac{M_p^2}{\mu^{2-m} \phi^m}\right| \ll 1 . 
 \label{qfg}
\end{equation}

Now we need approximations more precise than (\ref{slow-roll}) for the
inflaton dynamics. Dropping terms in (\ref{1}) and (\ref{3}) that are 
clearly smaller than order $g \frac{\sigma^2}{\mu^{2-m}
\phi^m}$ corrections to (\ref{slow-roll}), one can write down
\begin{equation}
 \ddot{\phi} + 3 H \dot{\phi} \left( 1 - f \frac{\sigma^2}{\mu^{2-m}
			       \phi^m}\right) 
 \simeq -V' \left\{1 + \left(1-\frac{m}{n}\right) g
	     \frac{\sigma^2}{\mu^{2-m} \phi^m}\right\},
 \label{11}
\end{equation}
\begin{equation}
 3 H^2 M_p^2 \simeq \frac{1}{2} \dot{\phi}^2 + V \left(1 + g
 \frac{\sigma^2}{ \mu^{2-m} \phi^m}\right) .
 \label{12}
\end{equation}
Hence the number of e-foldings $d\mathcal{N} = da/a$ is given by
\begin{equation}
 d \mathcal{N} = H dt  =  H \frac{d\phi}{\dot{\phi}} 
 \simeq -\frac{V\, d\phi}{M_p^2 V'} 
 \left\{1 + \left(\frac{m}{n}g - f\right) \frac{\sigma^2}{\mu^{2-m}
  \phi^m} + \mathcal{O}(\epsilon,\eta)\right\},
 \label{dN}
\end{equation}
where we have expressed terms $\frac{\dot{\phi}^2}{V} \times
\mathcal{O}(1)$, $\frac{\ddot{\phi}}{V''} \times \mathcal{O}(1)$ as
$\mathcal{O}(\epsilon, \eta)$. 
One should note that unless $m\neq 0$, the modulus correction to the
inflaton potential (i.e. the term with $g$) is canceled between
(\ref{11}) and (\ref{12}), and leaves no effect at linear order in
(\ref{dN}). 
Now, (\ref{slow-roll}) and
(\ref{sigma-g}) give 
\begin{equation}
 \frac{d \sigma}{d \phi} = \frac{\dot{\sigma}}{\dot{\phi}} \sim 2g
  \frac{V}{V'} \frac{\sigma}{\mu^{2-m} \phi^m},
 \label{a14}
\end{equation}
which is integrated to yield
\begin{equation}
 \ln \frac{\sigma}{\sigma_*} \sim 
  \left\{
  \begin{array}{cl}
  \dfrac{2g}{n} \ln \dfrac{\phi}{\phi_*} & (m=2) \\
   \dfrac{2g}{n(2-m)} \dfrac{\phi^{2-m} - \phi_*^{2-m}}{\mu^{2-m}}
  & (m\neq 2) \\
  \end{array}
 \right.
 \label{p16}
\end{equation}
where $\sigma_*$ and $\phi_*$ are their field values at some fixed
time. Substituting (\ref{p16}) into (\ref{dN}) yields an expression of
the form $d\mathcal{N} = F(\phi, \phi_*, \sigma_*)\, d\phi$. 
(However, note that the errors of the leading order approximation can
be amplified when taking the exponential of both sides of (\ref{p16})
before substitution. $|$(terms in (\ref{qfg})) $\times $ (\ref{p16})$|
\ll 1$ is required for good approximation.) 
By changing the variables $(\phi, \phi_*, \sigma_* ) \to (\tilde{\phi},
\phi, \sigma )$, the number of e-foldings obtained from a point
$(\phi,\sigma)$ in field space to the end of inflation is expressed as
\begin{equation}
 \mathcal{N}(\phi,\sigma) =  \int^{\phi_f}_{\phi}
  F(\tilde{\phi},\phi,\sigma)\,  d\tilde{\phi} ,
\end{equation}
where $\phi_f$ is the value of $\phi$ at the end of inflation. Then by
using the $\delta
\mathcal{N}$-formalism~\cite{Starobinsky:1986fxa,Sasaki:1995aw,Wands:2000dp,Lyth:2004gb},
the curvature perturbation is given by
\begin{equation}
 \delta \mathcal{N} = \mathcal{N}(\phi + \delta \phi,\sigma + \delta
  \sigma) -   \mathcal{N}(\phi, \sigma) =
 \frac{\partial \mathcal{N}}{\partial \phi} \delta \phi + 
 \frac{\partial \mathcal{N}}{\partial \sigma} \delta \sigma + \cdots.
\end{equation}

To leading order, the number of e-foldings obtained from $(\phi,
\sigma)$ is
\begin{equation}
 \mathcal{N}\sim \frac{\phi^2 - \phi_f^2}{2 n M_p^2}.
 \label{e-foldings}
\end{equation}
Since the $\sigma$-corrections are small, the surface where inflation
ends in field space is almost independent of the modulus field.
Therefore let us take $\phi_f$ as a constant for the moment. Later on,
we justify this assumption. Then, the part of $\delta \mathcal{N}$ linear
in $\delta \phi$ is given by 
\begin{equation}
 \frac{\partial \mathcal{N}}{\partial \phi} \sim
  \frac{V}{M_p^2 V'}  = 
  \frac{\phi}{n M_p^2}.
 \label{Nphi}
\end{equation}
This familiar result represents adiabatic curvature perturbations
generated due to $\delta \phi$ pushing patches of the universe
along the classical trajectories in the $\phi - \sigma$
plane.\footnote{Contribution to $\frac{\partial \mathcal{N}}{\partial
\phi}$ from the $\frac{\sigma^2}{\mu^{2-m} \phi^m}$ term in (\ref{dN})
represents differences among trajectories, similar to
$\frac{\partial \mathcal{N}}{\partial \sigma}$. This, compared to
(\ref{Nphi}), is suppressed by $g \frac{\sigma_f^2}{\mu^{2-m}
\phi^m}$. This factor can come close to or exceed unity only under $m<0$
and $\sigma_f^2 \gg M_p^2$.}

The leading contribution for $\partial \mathcal{N}/\partial \sigma$
comes from the second term in the curly brackets in (\ref{dN}). 
(Note that the $\mathcal{O}(\epsilon, \eta)$ terms may be larger than
the second term, but their dependence on $\sigma$ are schematically of
the form $\epsilon = \epsilon_0 \left(1 + \mathcal{O}\left(g
\frac{\sigma^2}{\mu^{2-m} \phi^m}\right) \right)$ with $\epsilon_0$ 
the value in the absence of modulus corrections, hence their contributions
to $\partial \mathcal{N}/ \partial \sigma$ are suppressed by
$\mathcal{O}(\epsilon, \eta)$.) Thus we obtain
\begin{equation}\label{Nsigma}
\begin{split}
 \frac{\partial \mathcal{N}}{\partial \sigma} 
 & \sim 
  \left\{
  \begin{array}{ll}
  \dfrac{1}{2} \left( \dfrac{2}{n} - \dfrac{f}{g} \right)
   \dfrac{\sigma}{M_p^2} 
 \left\{1 - \left(\dfrac{\phi_f}{\phi}\right)^{\frac{4}{n} g} \right\}
  & \quad (m=2) \\
  \dfrac{1}{2} \left( \dfrac{m}{n} - \dfrac{f}{g}\right)
   \dfrac{\sigma}{M_p^2} 
 \left\{1 - \exp \left(\dfrac{4g}{n (2-m) } \dfrac{\phi_f^{2-m} -
		  \phi^{2-m}}{\mu^{2-m}}\right)\right\} 
  & \quad (m\neq 2) \\
  \end{array}
 \right. \\
 & \sim \frac{1}{2} \left(\frac{m}{n} - \frac{f}{g} \right) 
 \frac{\sigma}{M_p^2} \left\{ 1 - \left(
  \frac{\sigma_f}{\sigma}\right)^2  \right\} .
\end{split}
\end{equation}
Note that we have used (\ref{p16}) in moving to the last
line.\footnote{When the $\sigma$-corrections are independent of $\phi$,
i.e. when $m=0$, our results (\ref{Nphi}) and (\ref{Nsigma}) can
be checked to agree with the result~(3.24) in
\cite{GarciaBellido:1995qq} where generic two-field inflation models
with separable potentials are studied.} This shows that when $\sigma$
has a positive effective mass squared (i.e. $g > 0$, provided $\phi$
and $\mu$ being positive) and $\sigma $ rolls towards zero, then
$\partial \mathcal{N}/\partial \sigma \propto \sigma /M_p^2$. On the
other hand, when the modulus has a tachyonic mass (i.e. $g < 0$) and
$|\sigma |$ grows, then adjacent trajectories diverge towards the end of
inflation and curvature perturbations are amplified by the factor
$-(\sigma_f / \sigma )^2$. Major conversion of entropy
perturbations from~$\delta \sigma$ to adiabatic curvature
perturbations~$\delta \mathcal{N}$ occurs at early (later) times for
converging (diverging) trajectories.

The ratio between linear contributions to $\delta \mathcal{N}$ from
$\delta \phi$~(\ref{Nphi}) and $\delta \sigma$~(\ref{Nsigma})
takes the form,
\begin{equation}
 \frac{\partial \mathcal{N} / \partial \sigma}{\partial \mathcal{N} /
  \partial \phi} = \mathcal{O}(1) \times
  \frac{\sigma}{ \phi} \left\{ 1 - \left(
				    \frac{\sigma_f}{\sigma}\right)^2
		       \right\}.
 \label{comp}
\end{equation}
Especially under $|\sigma| < | \phi | $ which is expected to hold for
most microscopic models (e.g., Section~\ref{sec:IV}), then curvature
perturbations from $\sigma$ can become dominant over that from $\phi$
only when the modulus has a tachyonic mass.
It is important to bear in mind that even when the absolute field value
of $\sigma$ is always smaller than that of $\phi$, a large growth of
$|\sigma |$ such that $\left| \frac{\sigma}{\phi} \left(
\frac{\sigma_f}{\sigma}\right)^2\right| \gtrsim 1$ makes contributions from
$\delta \sigma$ significant. 
(However one should also note that this ``amplification
factor''~$(\sigma_f / \sigma )^2$ is determined by $\phi$ and $\phi_f$,
cf. (\ref{Nsigma}).)
\\ 

Before turning to properties of curvature perturbations from $\delta
\sigma$, let us briefly discuss possible corrections to the above
formulae. First of all, where inflation ends (which is roughly when the
slow-roll parameters become of order unity, i.e. $\phi_f =
\mathcal{O}(1) \times M_p$) is not 
a constant but is slightly shifted by the modulus corrections. This can
be written schematically in the form
$\phi_f = \phi_{f_0} \left( 1 + \mathcal{O} \left( g
\frac{\sigma_f^2}{\mu^{2-m} \phi_f^m}\right) \right)$ with $\phi_{f_0}$
the value of $\phi_f$ in the absence of the $\sigma$-corrections. Hence
fluctuations of $\phi$ and $\sigma$ can modulate $\phi_f$, and curvature
perturbations can be generated at the end of
inflation~\cite{Lyth:2005qk,Alabidi:2006wa}.
Furthermore, we should also note that different
trajectories in the $\phi - \sigma$ plane end inflation at different
energy densities. This starts the subsequent evolution of the universe
from different temperatures, leading to extra $\delta \mathcal{N}$ after
inflation (see, e.g.~\cite{Sasaki:2008uc}). 
However, one can check that as long as (\ref{qfg}) holds during
inflation, both types of corrections are
subdominant, i.e., such corrections sourced by $\delta \phi$ and $\delta
\sigma$ are negligible compared to (\ref{Nphi}) and (\ref{Nsigma}),
respectively.\footnote{To be precise, only when $\sigma \sim \sigma_f$,
the additional $\delta \mathcal{N}$ sourced from $\delta \sigma$ can
become comparable to (\ref{Nsigma}). However in such cases,
curvature perturbations from $\delta \sigma$ are suppressed anyway
(cf. (\ref{comp})), and are negligible unless $ \sigma ^2 \gg \phi^2$.} 
We demonstrate the calculations only for the case 
we study in Subsection~\ref{subsec:gzero}, but extending the
estimations there ((\ref{deltaNf}), (\ref{deltaNc})) to the present case
is straightforward. 

We should also remark that in actual cases, the conditions~(\ref{qfg})
may break down during inflation. Inflation 
approaches an end when $M_p^2/\phi^2 \to \mathcal{O}(1)$, but can also
be terminated by the $\sigma$-correction~$ \left| g
\frac{\sigma^2}{\mu^{2-m}\phi^m} \right|$ approaching unity. Meanwhile,
inflation is not directly affected by $\left| g
\frac{M_p^2}{\mu^{2-m}\phi^m}\right| \to \mathcal{O}(1)$, but this will
accelerate/decelerate $\sigma$ and then the
approximation~(\ref{sigma-g}) becomes invalid. Here, we emphasize that
when $\sigma^2 \gg \sigma_f^2 $, i.e. the trajectories converge, most of
the conversion of entropy to curvature perturbations occur at early
times, hence the breakdown of (\ref{sigma-g}) at later times does not
affect the result~(\ref{Nsigma}) significantly.\footnote{However, as
discussed above, when (\ref{qfg}) breaks down then
additional~$\delta \mathcal{N}$ produced at the end and after inflation
may become non-negligible.} But when  $\sigma^2 \ll
\sigma_f^2 $, $\delta \mathcal{N}$ is generated substantially towards
the end of inflation and (\ref{qfg}) is required all through
in order to trust (\ref{Nsigma}).  \\

Now let us focus on the curvature perturbations~(\ref{Nsigma}) generated
by $\sigma$. By using $\mathcal{P}_{\delta \sigma}^{1/2} =
H/2 \pi$, the power spectrum is 
\begin{equation}
 \mathcal{P}_{\zeta_{\sigma}} = \left. \left( \frac{\partial
				    \mathcal{N}}{\partial 
			   \sigma}\right)^2  \left(\frac{H}{2
			   \pi}\right)^2
 \right|_{k = aH},
\end{equation}
where the right hand side is to be estimated at the time of horizon
crossing. Then its spectral index can be computed,
\begin{equation}
 n_s - 1 = \frac{d \ln \mathcal{P}_{\zeta_{\sigma}}}{d\ln k} \sim
  \frac{\sigma_f^2 + \sigma^2}{\sigma_f^2 - \sigma^2}\, 
 4 g \frac{M_p^2}{\mu^{2-m} \phi^m} 
 - n^2 \frac{M_p^2}{\phi^2},
\end{equation}
where one should recall that $\phi_f$ is fixed. In addition,
non-Gaussianity is sourced by the second derivative of $\mathcal{N}$
with respect to $\sigma$, 
\begin{equation}
 \frac{\partial^2 \mathcal{N}}{\partial \sigma^2}
 \sim \frac{1}{2} \left(\frac{m}{n} - \frac{f}{g} \right) 
 \frac{1}{M_p^2} \left\{ 1 - \left(
  \frac{\sigma_f}{\sigma}\right)^2  \right\}.
\end{equation}
Defining the non-Gaussianity parameter simply as the ratio between the
three-point function and the squared two-point function, we arrive at
\begin{equation}
 f_{\mathrm{NL}} = \frac{\langle \delta \mathcal{N}^3 \rangle}{\langle
  \delta \mathcal{N}^2 \rangle ^2} 
 = \frac{1}{2} 
 \frac{\partial ^2 \mathcal{N}/\partial \sigma^2}{(\partial \mathcal{N}
 / \partial \sigma )^2} \sim
 \left(\frac{m}{n} - \frac{f}{g}\right)^{-1} 
 \frac{M_p^2}{\sigma^2 - \sigma_f^2} .
 \label{fNL}
\end{equation}
One clearly sees that when $\sigma_f^2 \gg \sigma^2$ (which is required
for $\sigma$ with small field values to generate significant curvature
perturbations), then $f_{\mathrm{NL}} \propto M_p^2 / \sigma_f^2$ and a
sub-Planckian modulus field produces large non-Gaussianity. \\

Before ending this section, let us pause to note implications for
inflation models other than with $V(\phi) \propto \phi^n$. Given
other inflaton potentials, (\ref{a14}) is integrated to yield
results different from (\ref{p16}), leading to different growing or
damping rate of $| \sigma |$. If then the growing rate of a
tachyonic~$\sigma$ is large, as we emphasized in the discussions
above, $\partial \mathcal{N}/ \partial \sigma$ is significantly
amplified. We also mention that for small-field (i.e. $|\phi | \ll M_p
$) slow-roll models, $|\frac{V}{V' \phi}| = \frac{1}{\sqrt{2 \epsilon}} | 
\frac{M_p}{\phi} |$ is always larger than unity (in contrast to when
$V(\phi) \propto \phi^n$ where $= \mathcal{O}(1)$). Then for $m\neq
0$ cases, the leading modulation to the inflaton dynamics will be the 
$\frac{m V}{V' \phi} g \frac{\sigma^2}{\mu^{2-m} \phi^m}$ term in the
right hand side of (\ref{1}).

\section{Cases with Other Types of Modulus Corrections}
\label{sec:III}

In the previous section we have seen that light moduli with small field
values can leave their imprints on primordial curvature perturbations
only when they receive tachyonic backreaction and their field values are
amplified. However, this is not necessarily the case when
the modulus corrections enter the inflaton action in different
manners. The two interesting cases we study in this section are when the 
$\sigma$-correction only enters the inflaton kinetic term, and when the
$\sigma$-corrections in the inflaton kinetic term and potential have
different minima. 

\subsection{Modulating only the inflaton kinetic term}
\label{subsec:gzero}

First we study the case where the modulus correction only enters the
inflaton kinetic term, i.e. $g=0$ in (\ref{L}).
Here $\sigma$ modulates inflation without receiving significant
backreaction. 

For simplicity, we fix $m$ in (\ref{L}) to zero and study the
Lagrangian  
\begin{equation}
 \frac{\mathcal{L}}{\sqrt{-g}} = -\frac{1}{2} g^{\mu\nu} \partial_{\mu}
  \phi \partial_{\nu} \phi \left(1 - f \frac{\sigma^2}{\mu^{2}}\right) -
  \frac{1}{2} g^{\mu\nu} 
  \partial_{\mu} \sigma \partial_{\nu} \sigma - V(\phi).
\label{Lf}
\end{equation}
Let us again consider large-field inflation with $V (\phi) \propto \phi^n$,
where $n = \mathcal{O}(1)$. Here $\sigma$ receives an effective mass from
the inflaton kinetic term, therefore instead of (\ref{sigma-g}) the 
following approximation is assumed, 
\begin{equation}
 3 H \dot{\sigma} \sim -f \dot{\phi}^2 \frac{\sigma}{\mu^2}.
 \label{sigma-f}
\end{equation}
Then (\ref{slow-roll}) and (\ref{sigma-f}) require the necessary
conditions,
\begin{equation}
 \frac{M_p^2}{\phi^2}\ll 1 ,\qquad 
 \left| f \frac{\sigma^2}{\mu^2}\right| \ll 1,\qquad
 \left| \frac{M_p^2}{\phi^2} f \frac{M_p^2}{\mu^{2} }\right| \ll 1 ,
 \label{qf}
\end{equation}
where the last condition is equivalent to $m_{\sigma}^2 / H^2 \ll 1$. 
Now, (\ref{slow-roll}) and (\ref{sigma-f}) are combined to yield
\begin{equation}
 \ln \frac{\sigma}{\sigma_*} \sim \frac{n}{3} f \frac{M_p^2 }{\mu^2} \ln
  \frac{\phi}{\phi_*}.
\end{equation}
Since now the effective mass of $\sigma$ is suppressed by the slow-roll
parameter~$\epsilon$, $\sigma$ tends to roll less compared to the
previous section. Let us further assume
\begin{equation}
 \left| f \frac{M_p^2}{\mu^2} \right| \ll 1,
 \label{2-2}
\end{equation}
and study cases where the modulus field is fixed during inflation,
i.e. $\sigma \sim \mathrm{const.}$ Then from
\begin{equation}
 d \mathcal{N}  \simeq -\frac{V\, d\phi}{M_p^2 V'} 
 \left\{1 - f \frac{\sigma^2}{\mu^{2}} +
  \mathcal{O}(\epsilon,\eta)\right\},
 \label{dN-f}
\end{equation}
we arrive at
\begin{equation}
 \frac{\partial \mathcal{N}}{\partial \sigma } \sim - \frac{\phi^2 -
  \phi_f^2}{n M_p^2} f \frac{\sigma}{\mu^2} \sim -2 \mathcal{N} f
  \frac{\sigma}{\mu^2 },
 \label{Nsigma-f}
\end{equation}
where $\mathcal{N}$ is the number of e-foldings obtained between $\phi$
and $\phi_f$, cf. (\ref{e-foldings}). Since $\sigma \sim
\mathrm{const.}$, its correction to the inflaton dynamics stays constant
all through inflation and the resulting curvature perturbation becomes
proportional to $\mathcal{N}$. This means that the conversion of the
entropy to curvature perturbations happen more at larger $\phi$, where
more e-foldings are generated. 

Note that (\ref{Nsigma-f}) can also be obtained as a $ \sigma \to
\sigma_f$ limit of (\ref{Nsigma}) with $m=0$. (Remember that when
$m=0$, the $\sigma$-correction to the inflaton potential had no effect
on $\delta \mathcal{N}$, cf. (\ref{dN}).) However we emphasize that in
(\ref{Nsigma}), $\sigma \sim \sigma_f$ suppresses the curvature
perturbation to $| \partial \mathcal{N}/ \partial \sigma | \ll | \sigma /
M_p^2 |$. (Especially for $m=0$, $\left| g \frac{\phi^2}{\mu^2}\right|
\ll 1 $ was the condition for $\sigma \sim \sigma_f$.)
Whereas in the present case~(\ref{Nsigma-f}), one need not require
$\left| f \frac{\phi^2}{\mu^2}\right| \ll 1$ for $\sigma$ staying
constant, therefore $| \partial \mathcal{N} / \partial \sigma |$ can
become larger than $|\sigma / M_p^2 |$. This is nothing but stating that
$\sigma$ tends to roll less in the present case due to the small
effective mass induced by the inflaton kinetic term. 

Curvature perturbations from $\delta \phi$ is the same as in the
previous section, i.e. (\ref{Nphi}). Then the ratio between
contributions from $\delta \phi$~(\ref{Nphi}) 
and $\delta \sigma$~(\ref{Nsigma-f}) is now
\begin{equation}
 \frac{\partial \mathcal{N} / \partial \sigma}{\partial \mathcal{N} /
  \partial \phi} \sim - f \frac{\sigma \phi }{\mu^2},
\label{comp-f}
\end{equation}
where we have used $\phi^2 \gg \phi_f^2 = \mathcal{O} (1) \times
M_p^2$. 
For example, $\phi \sim 10 M_p$, $\sigma \sim M_p$, and 
$f\frac{M_p^2}{\mu^2} \sim \frac{1}{10}$ make curvature perturbations
from $\delta \phi$ and $\delta \sigma$ comparable.  \\

Before continuing, let us look into additional $\delta \mathcal{N}$
produced at the end of and after inflation, which was briefly discussed
in the previous section.

Where inflation ends is shifted as
$\phi_f = \phi_{f_0} \left( 1 + \mathcal{O} \left( f
\frac{\sigma^2}{\mu^2}\right)\right)$ where $\phi_{f_0}$ is the value in
the absence of the $\sigma$-correction, hence modulation of $\phi_f$ is
of order (note that when $\sigma $ stays constant, $\delta \phi$ does
not modulate $\phi_f$)
\begin{equation}
 \frac{\delta \phi_f }{\phi_f} = \mathcal{O} \left(f
	     \frac{\sigma}{\mu^2} \right) \delta \sigma.
\end{equation}
This modulates (\ref{e-foldings}) and yields
\begin{equation}
 \delta \mathcal{N}_f \sim - \frac{\phi_f}{n M_p^2} \delta \phi_f 
 =  -\frac{\phi_f^2}{n M_p^2} \mathcal{O} \left( f
					\frac{\sigma}{\mu^2}\right)
 \delta \sigma.
 \label{deltaNf}
\end{equation}
Since $\phi^2 \gg \phi_f^2$, 
this contribution at the end of inflation is small compared to 
(\ref{Nsigma-f}) which is mostly generated at large~$\phi$. 

Further $\delta \mathcal{N}$ can be produced after inflation, due to the
fact that the energy density is not equal on the surface where inflation
ends in field space. This effect can be estimated by computing the modulation
of the potential energy at the end of inflation,
\begin{equation}
 \delta \mathcal{N}_c = \frac{1}{4} \frac{\delta V_f}{V_f} =
  \frac{n}{4} \frac{\delta \phi_f}{\phi_f} = 
 \mathcal{O} \left( f \frac{\sigma}{\mu^2}\right) \delta \sigma ,
 \label{deltaNc}
\end{equation}
where we have assumed that the universe becomes radiation dominated
right after inflation ends. One sees that this effect is of the same
order as (\ref{deltaNf}), hence can also be ignored. \\

Focusing on the curvature perturbation sourced by $\delta
\sigma$~(\ref{Nsigma-f}), its spectral index (note especially
that $\mathcal{P}_{\zeta} \propto \mathcal{N}^2$ induces $n_s - 1
\supset -2/\mathcal{N}$) is
\begin{equation}
 n_s -1 \sim 2 \left(-\frac{1}{\mathcal{N}} + \frac{\dot{H}}{H^2}
     \right) \sim -n (n+4) \frac{M_p^2}{\phi^2} ,
\end{equation}
and the non-Gaussianity parameter defined as in (\ref{fNL}) is
\begin{equation}
 f_{\mathrm{NL}} \sim -\frac{1}{4 \mathcal{N}}\frac{\mu^2}{ f \sigma^2}
 \sim -\frac{n}{2} \frac{M_p^2}{\phi^2} \frac{\mu^2}{f \sigma^2} .
\end{equation}
\\

In this subsection we have mainly focused on the case where $\sigma$
stays constant during inflation, but extending the calculations to 
rolling $\sigma$ is straightforward. Basically, as one can guess
easily, $\partial \mathcal{N} / \partial \sigma$ is reduced (amplified)
for $\sigma$ with large positive (negative) $m_{\sigma}^2$. However, in
contrast to what we have studied in the previous section, one finds that
even in cases where $\sigma$ rolls towards zero, the curvature
perturbations sourced by sub-Planckian $\sigma$ can exceed 
that from super-Planckian $\phi$. 

\subsection{$\sigma$-corrections with different minima}
\label{subsec:h}

In actual cases, it may well be that the $\sigma$-corrections show up
both in the inflaton kinetic term and potential, but with different
minima. Then after redefining $\sigma$ so that the minimum of the
potential correction is zero (and also absorbing extra constant terms to
$\phi$ ), we arrive at the Lagrangian
\begin{equation}
 \frac{\mathcal{L}}{\sqrt{-g}} = -\frac{1}{2} g^{\mu\nu} \partial_{\mu}
  \phi \partial_{\nu} \phi \left(1 - h \frac{\sigma }{\mu} -
f \frac{\sigma^2}{\mu^{2}}\right) -
  \frac{1}{2} g^{\mu\nu} 
  \partial_{\mu} \sigma \partial_{\nu} \sigma 
 - V(\phi) \left( 1 + g \frac{\sigma^2}{\mu^2} \right) .
\label{Lh}
\end{equation}
Now the linear $\sigma$-correction (i.e. the term with $h$) gives
additional modulation. However for large $\sigma$, i.e. $|\sigma | \gg
\left| \frac{h}{f} \mu \right|$, the $h$-term becomes negligible
compared to the $f$-term. Then the difference in minima can be ignored
and the previous discussions are applied. 

On the other hand when $\sigma $ is small as $| \sigma | \ll \left|
\epsilon \frac{h}{g} \mu \right|$, then now the $g$-term becomes
negligible, i.e., the leading backreaction to $\sigma$, and modulation
to the inflaton dynamics are both sourced by the $\sigma$-correction in
the inflaton kinetic term. Thus we recover the situation in
Subsection~\ref{subsec:gzero}. \\

A behavior unique to (\ref{Lh}) arises when $\sigma $ is in
the intermediate range so that the $\sigma$-dynamics is
controlled by the $g$-term, while the leading
modulation to the inflaton dynamics is given by the linear $h$-term.

If $|g| \gtrsim |f|$, then the necessary conditions for
(\ref{slow-roll}) and 
\begin{equation}
 3 H \dot{\sigma} \sim -2 g V \frac{\sigma}{\mu^2} \label{40}
\end{equation}
to hold under the Lagrangian~(\ref{Lh}) are that the absolute values of
the following quantities to be smaller than one,
\begin{equation}
 \epsilon,\quad  \eta,\quad h \frac{\sigma}{\mu},\quad g
  \frac{\sigma^2}{\mu^2}, \quad g \frac{M_p^2}{\mu^2}, \quad \epsilon
  \frac{h}{g} \frac{\mu}{\sigma}.
\end{equation}
The last quantity is required to be small so that the $\sigma$-dynamics
(i.e. the right hand side of (\ref{40})) is given by the $g$-term,
not the $h$-term. (Some terms in (\ref{6}) are absent
here since we are considering the $m=0$ case in (\ref{L}) with an
additional linear $h$-term.) Then, computing the curvature perturbations
from $\delta \sigma$ as in Section~\ref{sec:II}, now we obtain an
additional contribution from the $h$-term,
\begin{equation}
 \frac{\partial \mathcal{N}}{\partial \sigma}
 \supset
 \frac{\partial}{\partial \sigma}\int^{\phi_f}_{\phi}
 \frac{\widetilde{V} d\tilde{\phi}}{M_p^2 \widetilde{V}'} h
 \frac{\tilde{\sigma}}{\mu} 
 \sim 
 - \frac{h}{2g} \frac{\mu}{M_p^2} \left( 1- \frac{\sigma_f}{\sigma
  }\right).
\end{equation}
Note that this result holds for arbitrary inflaton
potential~$V(\phi)$. This contribution becomes dominant over that from
the $f$-term: $\frac{\partial \mathcal{N}}{\partial \sigma }\supset
-\frac{f}{2g} \frac{\sigma}{M_p^2} \left\{ 1 - \left(
\frac{\sigma_f}{\sigma}\right)^2 \right\}$ (this $m=0$ result
from (\ref{Nsigma}) is also valid for arbitrary $V(\phi)$) when
$|\sigma + \sigma_f| \ll \left| \frac{h}{f} \mu \right|$. (Here, since
$m=0$, the $g$-term does not modulate~$d\mathcal{N}$ at its linear
order~$g \frac{\sigma^2}{\mu^2} $. However if $|g| \gg |f|$, then its
higher order contributions may exceed the linear order contribution from
the $f$-term.) Thus, $\sigma$ can source curvature perturbations in a
way different from the previous cases in the $\left| \epsilon \frac{h}{g}
\mu \right| \ll |\sigma | \ll \left|\frac{h}{f} \mu \right| $ region.

\section{Examples: Light Fields in D-Brane Inflation Models}
\label{sec:IV}

For concrete realizations of the above discussions, let us study D-brane
inflation models in string theory.
Usually in this class of models the position of a D-brane plays the role
of the inflaton, but its dynamics can be modulated by other physical
degrees of freedom of the brane.
In this section we focus on the extra spatial directions in the internal
manifold, and KK modes (oscillation modes) of wrapped branes.

\subsection{Monodromy-driven inflation with wrapped D-branes}
\label{subsec:KK}

When inflation is driven by D-branes wrapped on cycles of the internal
geometry~\cite{Kobayashi:2007hm,Becker:2007ui,Silverstein:2008sg},
KK modes in the wrapped direction(s) (i.e. oscillation modes) can
modulate the zero mode dynamics. 
In this subsection we focus on the model proposed in
\cite{Silverstein:2008sg} where monodromy elongates the wrapped cycle
and yields large-field inflation. In such case the KK modes become more
and more light as the cycle becomes 
large.\footnote{For a brane wrapped along a direction~$\lambda_1$ with
cycle length~$l_1$, the elongation of the wrapped cycle can be
understood as the monodromy allowing the brane to wrap multiple
times~$m$ along another direction~$\lambda_2$ (with length~$l_2$), while
spaced evenly in the $\lambda_1$ direction. Therefore the mass of the KK
modes is of order $m_{\mathrm{KK}} \sim 1/m l_2$, cf. $n^2/p^2$~term in
(\ref{A5}). Furthermore, when the spacing between the multiply wrapped
brane becomes narrow, light open string modes can also emerge with mass
$m_{\mathrm{open}} \sim l_1/m \alpha'$. However, we note that these open
strings stretching between the wrapped brane are basically heavier than
the KK modes as long as the length scales of the internal space are
larger than the string scale, i.e. $l_1 l_2 > \alpha'$. We thank
Hirosi Ooguri for helpful discussions on this point.}\\

In the paper~\cite{Silverstein:2008sg}, the authors consider
ten-dimensional type IIA string theory compactified on an orientifold of
a product of two nil three-manifolds~\cite{Silverstein:2007ac}. 
The nil three-manifold~$\mathcal{N}_3 $ has the geometry
\begin{equation}
\begin{split}
 \frac{ds_{\mathrm{nil}}^2}{\alpha'} 
 & = L_{u_1}^2 du_1^2 + L_{u_2}^2 du_2^2 + L_x^2
  \left( dx + \frac{M}{2} [u_1 du_2 - u_2 du_1]\right)^2 \\
 & = L_{u_1}^2 du_1^2 + L_{u_2}^2 du_2^2 + L_x^2 (dx' + M u_1 du_2)^2,
\end{split}
\end{equation}
(where $x' = x-\frac{M}{2} u_1 u_2$) compactified by 
\begin{equation}
\begin{split}
 (x, u_1, u_2) & \rightarrow (x+1, u_1, u_2), \\
 (x, u_1, u_2) & \rightarrow \left( x-\frac{M}{2} u_2, u_1+1, u_2
			      \right), \\
 (x, u_1, u_2) & \rightarrow \left( x+\frac{M}{2} u_1, u_1, u_2 +1
			      \right) .
\end{split}
\end{equation}
Then assuming a D4-brane wrapped along the $u_2$~direction and moving
along $u_1$ (with fixed $x'$), its position in the
$u_1$~direction can become the inflaton. 

Let us now derive the inflaton action obtained in 
\cite{Silverstein:2008sg}, but this time including the KK modes. 
We give in the appendix a general expression for the four-dimensional 
effective action obtained from a wrapped 4-brane. The formulae there are
applied to the present case by $u_1 \to r$ and $u_2 \to \lambda/2 \pi
$. Henceforth, we set $u_1 =r$ and  use $r$ instead of $u_1$. Then for
large~$r$, i.e. $r \gg L_{u_2} / M L_x$,
the relevant six-dimensional part of the metric takes the
form~(\ref{A1}) with
\begin{equation}
\begin{split}
 g_{rr} & = \alpha ' L_{u_1}^2 \equiv A^2,  \\
 g_{\lambda\lambda} & = \frac{1}{(2 \pi)^2}  \alpha ' L_x^2 M^2 r^2
 \equiv  B^2 r^2 .
 \label{ABr}
\end{split}
\end{equation}
Then for zero $B_2$ field in the world volume directions, and vanishing
world volume gauge field strength, one can compute the DBI
(or simply the Nambu-Goto) action of the D4-brane of the
form~(\ref{nambugoto}). After integrating out the wrapped direction, we 
obtain the four-dimensional effective action for the zero mode
position of the brane with an infinite tower of KK modes. 
(For detailed derivation, see the appendix.)
The zero mode~$r_0$ and the KK modes~$r_n$ ($n \neq 0$) are almost
canonically normalized by 
\begin{equation}
 \phi \equiv \frac{2}{3} A (2 \pi p T_4 B )^{1/2} r_0^{3/2}, \qquad
 \psi_n  \equiv  A (2 \pi p T_4 B )^{1/2} r_0^{1/2} r_n.
 \label{cannor}
\end{equation}
Considering small oscillations around the zero mode position, i.e.
$|r_n | \ll r_0 $, or $|\psi_n | \ll  \phi $, 
the action up to two derivatives and quadratic order in $(r_n,
\partial_\mu r_n)$ is obtained as follows,
\begin{equation}
\begin{split}
 S = \int d^4 x \sqrt{-g^{(4)}} 
 & \left[
-\frac{1}{2} (\partial \phi)^2 
 \left( 1 - \frac{2}{9} \frac{A^2}{B^2}
  \sum_{n\neq 0} \frac{n^2}{p^2 } \frac{|\psi_n |^2}{\phi^2} -
 \frac{1}{3}  \sum_{n\neq 
  0}\frac{|\psi_n |^2}{\phi^2}  \right) 
 - \frac{1}{2} \sum_{n\neq 0} (\partial \psi_n)(\partial
 \overline{\psi}_n) 
 \right.  \\
 & \quad \left. -  \left(\frac{3 \pi p T_4 B}{A}\right)^{2/3} \phi^{2/3}
 \left( 1 +  \frac{2}{9} \frac{A^2}{B^2}
  \sum_{n\neq 0} \frac{n^2}{p^2 } \frac{|\psi_n |^2}{\phi^2}  \right)
 - \frac{1}{3} \sum_{n\neq 0} \frac{\psi_n}{\phi } (\partial \phi)
 (\partial \overline{\psi}_n)
 \right],
 \label{KKS}
\end{split}
\end{equation}
where $p$ is the winding number. Dropping the KK modes~$\psi_n$, one
recovers the inflaton action obtained in \cite{Silverstein:2008sg} with
potential $V(\phi) \propto \phi^{2/3}$. 
Here, the $\psi_n$-correction to the inflaton kinetic term can be
understood as nonzero $\psi_n$ increasing the wrapped volume, therefore
correcting the canonical normalization~(\ref{cannor}). The correction to
the inflaton potential is due to the brane tension damping the
oscillation, which gives $\psi_n$ effective mass which becomes lighter
for larger~$\phi$, i.e. for longer cycle. 
However, the KK modes also receive tachyonic instability
from their coupling to the zero mode motion (note that during inflation
$a^3 \sum \frac{\psi_n}{\phi } \dot{\phi} \dot{\overline{\psi}_n} \simeq
-\frac{3}{2} a^3  H \frac{\dot{\phi}}{\phi} \sum | \psi_n |^2 +
(\mathrm{total\, \, derivatives})$). This effective tachyonic mass
can be seen explicitly if instead of $\phi$ we introduce
\begin{equation}
  \varphi \equiv \frac{2}{3} A (2 \pi p T_4 B )^{1/2} r_0^{3/2}
 \left( 1 + \frac{3}{8} \sum_{n\neq 0} \frac{|r_n |^2}{r_0^2} \right),
\end{equation}
for which the action becomes
\begin{equation}
\begin{split}
 S = \int d^4x \sqrt{-g^{(4)}} 
 & \left[ -\frac{1}{2} (\partial \varphi)^2 
 \left( 1 - \frac{2}{9} \frac{A^2}{B^2}
  \sum_{n\neq 0} \frac{n^2}{p^2 } \frac{|\psi_n |^2}{\varphi^2} \right) 
 - \frac{1}{2} \sum_{n\neq 0} (\partial \psi_n)(\partial
 \overline{\psi}_n) \right. \\
 & \quad \left. -  \left(\frac{3 \pi p T_4 B}{A}\right)^{2/3} \varphi^{2/3}
 \left( 1 +  \frac{2}{9}\frac{A^2}{B^2}
  \sum_{n\neq 0}  \frac{n^2}{p^2 } \frac{|\psi_n |^2}{\varphi^2}
  - \frac{1}{9}  \sum_{n\neq 0}\frac{|\psi_n |^2}{\varphi^2} 
  \right) \right] .
 \label{KKS2}
 \end{split}
\end{equation}
Basically, if $A^2 \ll B^2$ the lower KK modes obtain tachyonic
instability. (One can image a cone which spreads out quickly away from
its tip.) 

We comment that, as was discussed in \cite{Silverstein:2008sg}, the
$\mathcal{N}_3 \times \widetilde{\mathcal{N}}_3$ space also allows a 
variant inflaton action. Given a different configuration for
the D4-brane (e.g. by wrapping the brane along the $u_2 - \tilde{u}_2$
direction while moving it in a linear combination of $u_1 + \tilde{u}_1$
and $u_2 + \tilde{u}_2$ directions with $u_1 = \tilde{u}_1$), a metric 
$g_{rr} \propto r^2$, $g_{\lambda \lambda} \propto r^2$ can be 
realized for large~$r$, which gives an inflaton potential $V(\phi) \propto
\phi^{2/5}$. One can check that in this case, the $\psi_n$-correction
terms proportional to $\frac{n^2}{p^2}$ take the form $\frac{|\psi_n 
|^2}{\mu^{4/5} \phi^{6/5}}$. \\

Going back to the action (\ref{KKS}) or (\ref{KKS2}), 
the example values chosen in \cite{Silverstein:2008sg} for the
parameters give $\frac{A^2}{B^2} = (2 \pi)^2  \frac{L_{u_1}^2}{L_x^2
M^2} \gtrsim 10^4$, so for winding number $p = \mathcal{O}(1)$ the
tachyonic instability is negligible. There, since the CMB scale
leaves the horizon at $\phi/M_p = \mathcal{O} (10)$ for $V(\phi) \propto
\phi^{2/3}$, the effective mass of $\psi_n$ becomes heavier than the
Hubble parameter, i.e. the last condition of (\ref{qfg}) is not
satisfied. However, since the parameters can be shifted in variants of
the construction, let us give general discussions for cases where the
metric takes the form~(\ref{ABr}). 

We assume $\frac{A^2}{B^2} \frac{n^2}{p^2} \ll 10^3$, so that the
effective mass of $\psi_n$ is lighter than the Hubble parameter when the
CMB scale exits the horizon. Then, for $\frac{A^2}{B^2}
\frac{n^2}{p^2} \gg 1$, the KK modes~$\psi_n$ 
roll towards zero and from the discussions given in
Section~\ref{sec:II}, curvature perturbations generated from 
fluctuations of $\psi_n$ is negligible compared to that from the
inflaton~$\phi$. 
Meanwhile, if $\frac{A^2}{B^2} \frac{n^2}{p^2} \ll 1$ the KK
modes~$\psi_n$ are tachyonic. Substituting $n=2/3$, $m=2$, $f=0$, and
$g=-1/9$ to (\ref{Nsigma}), one finds the ``amplification factor''
(i.e. $(\sigma_f / \sigma )^2$) to be $(\phi_f / 
\phi)^{2/3} $, which is smaller than~10.

One can also check that when $\delta \phi$ sources the dominant
contribution to the curvature perturbations, non-Gaussianity produced by
the KK modes~$\psi_n$ is negligible, i.e. $\left| \frac{\partial^2
\mathcal{N}}{\partial \sigma^2}\left( 
\frac{\partial \mathcal{N}}{\partial \sigma}\right)^2  \Big/ 
\left( \frac{\partial \mathcal{N}}{\partial \phi}\right)^4 \right| \ll
1$,  for both $\frac{A^2}{B^2} \frac{n^2}{p^2} \gg 1$ and $\ll 1$ cases 
as long as $| \psi_n | \ll |\phi |$ holds during inflation. 

Therefore, for the metric~(\ref{ABr}) with
$\frac{A^2}{B^2} \frac{n^2}{p^2} \gg 1$ or $ \ll 1$, 
we conclude that unless the KK modes start off from Planckian or
super-Planckian values, we can safely neglect such modes during inflation.
However, one may also need to look into their cosmological implications
after inflation, since even though they become heavy after (or towards
the end of) inflation, they can carry on non-negligible oscillation
amplitudes from the early times in the inflationary era when they were
light. \\

Before ending this section, we note that given additional contributions
to the inflaton action other than from the Nambu-Goto part of the
DBI action, then the KK modes' corrections to the inflaton
potential can be weakened while the corrections to the inflaton kinetic
term remains effective. Especially for wrapped D-brane inflation models
in warped throats~\cite{Kobayashi:2007hm,Becker:2007ui} where one
expects various sources contributing to the inflaton potential, the cases
studied in Section~\ref{sec:III} may be realized and then one needs to
take into account of the KK modes seriously.

\subsection{Warped D-brane inflation}
\label{subsec:angular}

D3-brane inflation models in a warped throat
geometry~\cite{Kachru:2003sx,Baumann:2006th,DeWolfe:2007hd,Baumann:2007np,Krause:2007jk,Baumann:2007ah,Baumann:2008kq,Baumann:2010sx}
admit light field corrections from angular directions.
One may expect that a finite region of the throat is well modeled by a
noncompact throat (such as a conifold) possessing angular isometries,
plus corrections which break the isometries due to the throat being
glued to a compact bulk. Then, 
given that such corrections are well suppressed by the warping, light
angular directions can emerge in the regime deep into the throat. 
(Hence for an inflaton D3 moving towards the throat tip, light angular
directions are expected to show up at later times in the inflationary 
era, which is in contrast to the previous subsection where the KK modes
were lighter at earlier times. If inflation ends by brane annihilation,
then one may not have to worry about the cosmological moduli problem for
the light directions at all.)
Usually the model is treated as a single field model of the radial
position of the D3-brane, by fixing its angular positions at their local
minima and integrating out the angular degrees of freedom. However, in
cases where the angular directions become light, one may have to treat
such directions as dynamical fields.\footnote{Generating curvature
perturbations at the end of inflation (by shifting~$\phi_f$)
with light angular directions were
investigated in \cite{Lyth:2006nx,Chen:2008ada}. Also, in
\cite{Panda:2007ie}, the two-field dynamics of the D3's radial position
and the volume modulus was studied. For DBI inflation with multiple
fields, see e.g. \cite{Langlois:2008wt,Langlois:2008qf,Arroja:2008yy}.}

In simple cases, the D3-brane's radial~$r$ and periodic angular
directions~$\theta^i$ are canonically normalized to $\phi \propto r$ and 
$\sigma^i \propto r \theta^i$ (hence $|\phi| \gtrsim |\sigma^i |$). Then
one expects an inflaton potential corrected by terms~$\sigma^i/\phi$
multiplied by some powers of the warp factor.
Furthermore, if inflation ends by D3-$\overline{\mathrm{D}3}$
annihilation at the throat tip, then the
$\overline{\mathrm{D}3}$-brane's angular degrees of
freedom~$\tilde{\sigma}$ may also become light and give 
corrections of the form~$\tilde{\sigma}/\mu$. 

One can imagine that various contributions to the inflaton potential
have different dependence on angular positions of
D3/$\overline{\mathrm{D}3}$, 
giving an action highly more complicated than (\ref{L}). 
However, from the discussions in the previous sections, we expect that
the multi-field effects can become important if the
D3/$\overline{\mathrm{D}3}$ starts off near a local maximum of the
angular directions and rolls down during inflation. It would be interesting
to examine the effects of angular directions in a concrete and
computable setup of warped D3-brane inflation.

\section{Conclusions}
\label{sec:conc}

We have studied effects of light fields showing up as small corrections
in the inflaton action. While minimally affecting the inflaton dynamics,
such moduli can still leave imprints on cosmological observables through
generating curvature perturbations.\footnote{While in this paper we have
studied effects of light fields minimally affecting the inflaton
dynamics, heavy fields which considerably alter the inflaton dynamics
may also have interesting consequences. For example,
\cite{Tolley:2009fg} demonstrated cases where such heavy field effects
generate equilateral non-Gaussianity.} The fluctuations of the moduli are 
converted to curvature perturbations during inflation, due to patches of
the universe taking different trajectories in field space. The basic
picture is simple: for converging trajectories the transformation of
entropy to adiabatic perturbations is suppressed at later times, and
for diverging trajectories it is enhanced. The curvature perturbation
produced by the moduli is especially amplified and can dominate over
that from the inflaton when the moduli receive tachyonic backreaction
from the inflaton potential, or when the main moduli corrections enter
the inflaton kinetic term. In this paper we mainly focused on
modulating large-field inflation with $V(\phi) \propto \phi^n$, but it
would be interesting to perform a systematic study with more general
inflaton potentials. The effects of moduli can be simply stated as
modulating the relation between $d\mathcal{N}$ and $d \phi$, hence they
can operate in basically any inflation model. Modulating models beyond
the slow-roll ones may have distinct features and is worthy of investigation. 
(For e.g., rapid-roll~\cite{Linde:2001ae,Kofman:2007tr,Kobayashi:2009nv}
and DBI inflation
models~\cite{Silverstein:2003hf,Alishahiha:2004eh}. Both  
are well-motivated by the warped D-brane inflation setup, which 
provides many candidate light fields as was discussed in
Section~\ref{sec:IV}.) 

Our study has a range of applications. 
As in the D-brane inflation models we have studied in the paper, it is
important to examine the (in)validity of integrating out extra degrees
of freedom in inflation models consisting of multiple fields. 
Also, when discussing cosmological scenarios where the presence of
light fields are inevitable or required (e.g. curvaton, modulated
reheating scenario), one needs to look into the fields' effects not just
after inflation, but also during inflation. 
In this work, for some simple cases we provided conditions under which
the moduli can be ignored. On the other hand we also exhibited cases
where the light moduli dominantly source the curvature perturbations.
The mechanism of generating curvature perturbations from light moduli
instead of the inflaton may open up new possibilities for inflation
model building.

\begin{acknowledgments}
 We are grateful to Damien Easson, Masahiro Kawasaki, Tatsuma Nishioka,
 Hirosi Ooguri, and David Wands for useful conversations. The work of
 T.K. was supported by Grant-in-Aid for JSPS Fellows
 No.~21$\cdot$8966. The work of S.M. was supported in part by
 Grant-in-Aid for Young Scientists (B) No. 17740134, Grant-in-Aid for
 Creative Scientific Research No. 19GS0219, Grant-in-Aid for Scientific
 Research on Innovative Areas No. 21111006, Grant-in-Aid for Scientific
 Research (C) No. 21540278, and the Mitsubishi Foundation. This work was
 supported by World Premier International Research Center Initiative
 (WPI Initiative), MEXT, Japan. 
\end{acknowledgments}

\appendix

\section{Effective Action of a Wrapped 4-Brane}
\label{sec:app}

Here we give a general expression for the four-dimensional effective
action derived from the Nambu-Goto action of a 4-brane wrapping a
1-cycle. We consider the 4-brane to be stretching along the external
space~$x^\mu$ ($\mu = 0,1,2,3$), while wrapping the $\lambda$~direction
with winding number~$p$ and moving along the 
$r$~direction in the internal space. The six-dimensional part of the
metric which is relevant for us is assumed to take the following form,
\begin{equation}
 ds^2 = g^{(4)}_{\mu\nu} (x)\,  dx^\mu dx^\nu + g_{rr}(r)\,  dr^2 +
  g_{\lambda\lambda} (r)\,  d\lambda^2, 
 \label{A1}
\end{equation}
with the $\lambda$~direction compactified by $ \lambda \simeq
\lambda + 2 \pi$. Taking the brane coordinates to
coincide with $x^\mu$ and $\lambda $, the Nambu-Goto action of the 4-brane
is 
\begin{equation}
 S = -T_4 \int d^4 x \int_0^{2 \pi p} d\lambda\,   \sqrt{-\det
  \left(G_{MN}\,  \partial_m X^M    \partial_n X^N \right) },
 \label{nambugoto}
\end{equation}
(where $T_4$ is the 4-brane tension) with 
\begin{equation}
 \det \left(G_{MN}\, \partial_m X^M \partial_n X^N \right) = 
  g^{(4)} \left\{ g_{\lambda \lambda } +
       g_{\lambda  \lambda } g_{rr} g^{(4)\mu\nu}\partial_{\mu}r
       \partial_{\nu}r 
       + g_{rr} (\partial_{\lambda} r)^2 \right\},
\end{equation}
where $g^{(4)} = \det (g^{(4)}_{\mu\nu})$.

Now let us expand the radial position of the 4-brane as
\begin{equation}
 r(x^\mu, \lambda) = \sum_{n=-\infty}^{\infty} r_n (x^\mu)\, 
  e^{i \frac{n}{p}\lambda}
\end{equation}
where $\bar{r}_n = r_{-n}$.
Upon expanding the action~(\ref{nambugoto}) up to 
two $x^\mu$-derivatives and quadratic order in $(r_n,\,  \partial_\mu r_n)$, 
and then integrating out the $\lambda$~direction, we arrive at 
\begin{multline}
 S \simeq  -2 \pi p T_4 \int d^4 x \sqrt{-g^{(4)}}
 \sqrt{g_{\lambda\lambda}} \\ 
 \quad \times \Biggl[
  \frac{1}{2} g_{rr}g^{(4)\mu\nu} \partial_{\mu}r_0 \partial_{\nu}r_0 
  \left\{
1+\frac{1}{2} \sum_{n\neq 0} |r_n|^2 
 \left( \frac{1}{2} \frac{g_{\lambda\lambda}''}{g_{\lambda\lambda}} -
  \frac{1}{4} \left(
	       \frac{g_{\lambda\lambda}'}{g_{\lambda\lambda}}\right)^2 +
  \frac{g_{rr}' g_{\lambda\lambda}'}{g_{rr} g_{\lambda\lambda}} +
  \frac{g_{rr}''}{g_{rr}} - \frac{n^2}{p^2 }\frac{g_{rr}}{g_{\lambda\lambda}}
\right)
\right\} \\
  +\frac{1}{2} g_{rr}\sum_{n\neq 0} g^{(4)\mu\nu}
 \partial_{\mu} r_n  \partial_{\nu} \bar{r}_n
+\left(\frac{1}{2} \frac{g_{\lambda\lambda}'}{g_{\lambda\lambda}} +
  \frac{g_{rr}'}{g_{rr}}\right) g_{rr} \sum_{n\neq 0}r_n g^{(4)\mu\nu}
\partial_{\mu} r_0 \partial_{\nu} \bar{r}_n \\
 +1 
 +\frac{1}{2} \sum_{n\neq 0}|r_n|^2 
 \left(\frac{1}{2} \frac{g_{\lambda\lambda}''}{g_{\lambda\lambda}} -
  \frac{1}{4} \left(
	       \frac{g_{\lambda\lambda}'}{g_{\lambda\lambda}}\right)^2 +
 \frac{n^2}{p^2}\frac{g_{rr}}{g_{\lambda\lambda}} \right) \Biggr] , 
 \label{A5}
\end{multline}
where $\sum_{n\neq 0} \equiv \sum^{\infty}_{n=-\infty, \, n\neq 0}$.
The primes denote derivatives, and $g_{rr}$, $g_{\lambda\lambda}$ are
functions of the zero mode $r_0(x^{\mu})$.

%%%%%%%%%%%%%%%%%%%%%%%%%%%%%%%%%%%%%%%%%%%%%%%
%%%%%%%%%%%%%%%%%%%%%%%%%%%%%%%%%%%%%%%%%%%%%%%

\end{document}